\documentclass[conference]{IEEEtran}
\usepackage{graphicx}
\usepackage{xcolor}
\usepackage{tcolorbox}
\usepackage{amsmath}
\usepackage{pgfplots}
\usepackage{pgfplotstable}
\usepackage{pgfkeys}
\pgfplotsset{compat=1.18}
\usepackage{booktabs}
\usepackage{lscape}
\usepackage{makecell}
\usepackage{array}
\usepackage{float}
\usepackage{lmodern}
\usepackage{url}
\usepackage{enumitem}
\usepackage{xparse}
\usepackage{multirow}
\usepackage{multicol}
\usepackage{hyperref}
\usepackage{comment}

\definecolor{GreenYellow}{rgb}{0.8, 0.8, 0}

\newcommand{\linebreakand}{%
  \end{@IEEEauthorhalign}
  \hfill\mbox{}\par
  \mbox{}\hfill\begin{@IEEEauthorhalign}
}
\makeatother

\begin{document}

\title{Exploring the Use of LLMs for Requirements Specification in an IT Consulting Company}

\author{
    \IEEEauthorblockN{Liliana Pasquale}
    \IEEEauthorblockA{
        School of Computer Science\\
        University College Dublin, Dublin, Ireland \\
        liliana.pasquale@ucd.ie
    }
    \and   
    \IEEEauthorblockN{Azzurra Ragone}
    \IEEEauthorblockA{
        Department of Computer Science\\
        University of Bari "A. Moro", Bari, Italy \\
        azzurra.ragone@uniba.it
    }
    \and
    \IEEEauthorblockN{Emanuele Piemontese\textsuperscript{\textsection}}
    \IEEEauthorblockA{
        Department of Computer Science\\
        University of Bari "A. Moro", Bari, Italy \\
        e.piemontese@studenti.uniba.it
    }
    \and
    \IEEEauthorblockN{Armin Amiri Darban}
    \IEEEauthorblockA{
        Computer Engineering Department\\
        Polytechnic of Bari, Bari, Italy \\
        a.amiridarban@studenti.poliba.it
    }
}

\maketitle
\begingroup\renewcommand\thefootnote{\textsection}
\footnotetext{This author conducted a student internship at Fincons group.}
\endgroup

\begin{abstract}
In practice, requirements specification remains a critical challenge. The knowledge necessary to generate a specification can often be fragmented across diverse sources (e.g., meeting minutes, emails, and high-level product descriptions), making the process cumbersome and time-consuming. In this paper, we report our experience using large language models (LLMs) in an IT consulting company to automate the requirements specification process.
In this company, requirements are specified using a Functional Design Specification (FDS), a document that outlines the functional requirements and features of a system, application, or process. We provide LLMs with a summary of the requirements elicitation documents and FDS templates, prompting them to generate Epic FDS (including high-level product descriptions) and user stories, which are subsequently compiled into a complete FDS document. We compared the correctness and quality of the FDS generated by three state-of-the-art LLMs against those produced by human analysts.
Our results show that LLMs can help automate and standardize the requirements specification, reducing time and human effort. However, the quality of  LLM-generated FDS highly depends on inputs and often requires human revision. Thus, we advocate for a synergistic approach in which an LLM serves as an effective drafting tool while human analysts provide the critical contextual and technical oversight necessary for high-quality requirements engineering (RE) documentation.
\end{abstract}

\section{Introduction}
The requirements specification process remains a significant bottleneck and critical challenge across many industries. In practice, the knowledge needed to define requirements is often extensive~\cite{Bjarnason.2011}  and scattered across various sources~\cite{Inayat.2015}, such as meeting minutes, email correspondence, high-level product descriptions, and other elicitation artifacts. As a result, requirements analysts need to examine fragmented and large sets of information to identify and specify the system requirements, which can be cumbersome and time-consuming. This challenge highlights the need for automated approaches to streamline the requirements specification process. 

Large Language Models (LLMs) have the potential to address this challenge. They can quickly process and analyze lengthy, diverse and unstructured natural language content to extract system requirements specified in a pre-defined format~\cite{Arora.2024}. In this paper, we report our experience on using LLMs to automate the specification of requirements in an international IT consulting company, focusing on a project related to digital rights management. The requirements are documented using a custom template-Functional Design Specification (FDS)- which outlines the system's functional requirements, features, and behavior.

We explore the feasibility and effectiveness of utilizing LLMs to automate the generation of FDS documents,  aiming to reduce manual effort while preserving the quality of the specification. 

Additionally, We investigated the practical application of three state-of-the-art LLMs (specifically GPT-3.5-turbo, GPT-4o, and Llama-3-70B) within the requirements specification workflow of the company. The methodology involved providing these LLMs with summarized requirements elicitation documents (such as presentations and meeting minutes) and predefined FDS templates. The LLMs were prompted to generate key components of the FDS, including Epic FDS providing high-level product descriptions and user stories, which were subsequently compiled into complete FDS documents. 

To evaluate this approach, we systematically compared the correctness, quality, and structure of the LLM-generated FDS documents against those produced during the traditional, analyst-driven requirements specification process used within the company. We assessed the plausibility of the created user stories by evaluating whether they semantically covered those crafted by the analysts. In addition, We also qualitatively examined the structure and content of the entire FDS.
We assessed the syntactic and semantic qualities of the user stories using the Quality User Stories framework~\cite{Lucassen.RE.2015}, and qualitative feedback obtained by interviewing an experienced requirements analyst involved in the project. 

Our findings indicate that LLMs can automate and standardize the requirements documentation, offering tangible time savings compare to the manual effort typically required. However, the study also highlights critical limitations. The completeness and precision of LLM-generated requirements highly depend on the quality and comprehensiveness of the input information and often necessitate human review, refinement, and contextual information. This research contributes empirical insights into the practical application of LLMs in industrial requirements engineering (RE) settings, advocating for a synergistic human-AI collaboration where LLMs act as efficient drafting assistants, augmented by the indispensable contextual understanding and critical oversight provided by human analysts to ensure high-quality, reliable requirements documentation. 

The remainder of the paper is organized as follows: Section \ref{sec:related} reviews relevant related work on LLM-powered approaches in requirements engineering. Section \ref{sec:context} describes the industrial context of the study, while Section \ref{sec:setup} delves into the experimental setup, designed to replicate the company's requirements specification workflow. 
Section \ref{sec:eval} presents the evaluation of the quality of the LLM-generated artifacts. Section \ref{sec:discussion} discusses key findings and limitations. Finally, Section \ref{sec:conclusion} concludes the paper.

\section{Related Work}
\label{sec:related}

LLMs hold promise for improving the efficiency and accuracy of requirements-related tasks. Arora et al.~\cite{Arora.2024} envisioned an LLM-driven RE process in which LLMs automate different RE activities (elicitation, specification, analysis and validation). They claim that LLMs can streamline the specification process by converting the unstructured material collected during the elicitation stage into structured templates like user stories. 

Recent work has leveraged deep learning models and LLMs to automate the elicitation of software requirements. Gudaparthi et al.~\cite{Gudaparthi.RE.2023} proposed a novel approach that uses adversarial examples to identify new requirement candidates. By applying small perturbations to existing requirements via a deep neural network, their method generates traceable variations that facilitate stakeholder discussions while maintaining transparency to the original features. Wei et al.~\cite{Malej.ASE.2024} employed LLMs for app feature elicitation, using one approach inspired by LLMs and another based on existing app descriptions from app stores. Their results show that while both methods produce relevant and clearly described features, LLMs can handle novel, previously unseen app scopes. However, some features produced by LLMs may be unrealistic or unclear regarding feasibility, requiring human analysis. 

Bencheikh and Hoglund~\cite{Bencheikh.2023} conducted a study exploring the efficacy of AI tools, particularly ChatGPT, in efficiently generating software requirements. They compared ChatGPT-generated requirements with human-authored ones under identical domain descriptions. ChatGPT matched human performance in both functional and non-functional requirements without additional context, while offering superior time efficiency. Although this study acknowledged the time efficiency of ChatGPT, it also recognized that experienced human participants tend to produce more comprehensive requirements. Similarly, Ronanki et al.~\cite{Ronanki.SEEA.2023} used a set of standardized questions to elicit requirements from both ChatGPT and five RE experts. The LLM-generated requirements were more abstract, atomic, and consistent, though their practicality in real-world prioritization scenarios remains an open question. Zhang et al.~\cite{Zhang.ICASD.2024} explore the use of LLMs to enhance the quality of user stories in agile software development. In particular, they used LLMs to automate the refinement of user stories, demonstrating improvements in clarity, completeness, and alignment with business goals. While these enhancements were generally well-received, concerns were raised regarding the increased complexity and length of some user stories, particularly with the GPT-4 model. The study highlights the importance of human oversight to ensure that AI-generated stories remain relevant and accurate. 

Similarly to our study, Rahman et al.~\cite{Rahman.ICSME.2024} explore the use of LLMs to generate user stories from requirements documents. They propose the Refine and Thought (RaT) methodology to reduce the hallucination of LLMs when analyzing complex requirements documents. It is a refinement of the Chain of Thought (CoT) prompting technique, which guides the model through a step-by-step reasoning process. By shortening and sharpening the input, RaT ensures that the model can focus more effectively on the essential information, producing more consistent and reliable results, especially when dealing with lengthy and complex requirements documents. However, differently from our work, the authors have not used LLMs in a real industrial context nor compared the LLMs' performance with a ground truth set of requirements. Beyond requirements specification, LLMs have also been used in industry settings to support safety analysis in autonomous vehicles~\cite{Nouri.RE.2024}. They could identify hazardous events and derive associated safety goals more efficiently than manual methods.

\section{Company-Specific Requirements Specification Process}\label{sec:context}
This section outlines the context of the study, including the requirements specification processes, tools, and document formats employed by the Fincons Group\footnote{Fincons Group is an international IT consultancy founded in 1983 specializing in digital transformation. As of recent estimates, Fincons Group employs approximately 2,000 professionals worldwide, particularly in Italy, Switzerland, Germany, France, the UK, Belgium, the USA, and India. \textbf{The data used in this study is not fully publicly available due to privacy and confidentiality reasons.}}  - the company where the study was conducted.

We collected the data used in this study from the \textit{AllRights} project, which supports the management of digital rights across multiple domains and helps companies improve the efficiency and monetization of media content. 
As \textit{AllRights} is a complex and constantly evolving project, this study focused on a selected area of the entire application, characterized by intermediate complexity, ideal for experimentation: the ‘Availability’ functionality. 
The Availability functionality is designed to manage the distribution of digital content (e.g., movies) across various platforms and territories. Through a very intuitive interface, users can monitor in real-time when and where content can be distributed or broadcast while respecting license agreements. ‘Availability’ also allows users to configure alerts for the expiration of distribution time windows and integrates with other modules of the \textit{AllRights} platform, such as contract management and reporting.

Before initiating the  study, we analyzed the processes and tools used at Fincons to specify and manage the requirements.

The company uses two project management tools: Confluence\footnote{\url{https://www.atlassian.com/software/confluence}} and Jira\footnote{\url{https://www.atlassian.com/software/jira}}. 
The outcome of the requirements specification is the \textbf{FDS}, a document that outlines the intended functionalities and behaviors of the application. It acts as a blueprint that answers critical questions such as what the system should do, how different components interact, and how user needs will be fulfilled. While it does not dive into the implementation details or technical breakdown of each component, it provides a comprehensive overview of the system's logic and structure from a functional standpoint. The primary goal of an FDS is to bridge the gap between non-technical stakeholders who define the requirements and the development team, who must translate those requirements into working code. It ensures that both sides share a common understanding of the system’s goals, functionality, and scope, even if they speak entirely different  languages. 

This document also contains a link to the \textbf{User Stories} reported in Jira. The format for the user stories is the classical "Role-Goal-Benefit", using the typical structure: \textit{"As a... I want to... So that..."}.

Before filling out the FDS template in Confluence, the IT analysts at Fincons fill in two different templates that are used as input for the FDS compilation process: Epic FDS (stored in Confluence) and User Stories (US) document (stored in Jira). 
The \textbf{EPIC FDS} document includes information about objectives, constraints and assumptions, the high-level process flow, and US traceability Matrix. 
For a comprehensive description of the fields included in the FDS, Epic FDS and US document, please refer to the online appendix\cite{appendix2025}.

\section{Experimental Setup}\label{sec:setup}
This section outlines the key components of the experimental setup. This preliminary work aims to explore how LLMs can be used to automate the creation of documentation, providing crucial support in defining and organizing requirements and helping analysts reduce their workload. 
The objective is to assess whether these technologies can not only simplify the documentation process but also ensure the quality in terms of the consistency and comprehensibility of information crucial for software development.

The objective of our study was to understand how much the LLMs can help the company, and in particular the analysts, automate and speed up the requirement specification process. We used the LLMs to mimic the activities performed by requirements analysts during the specification process.

We selected GPT3.5-turbo (1106), GPT-4o, and Llama-3-70B-Instruct for the experiment. Hereafter, we refer to them as \textbf{GPT3}, \textbf{GPT4}, and \textbf{Llama}, respectively. Privacy and security policies of the company informed the selection of the models\footnote{We underline that the study was performed between June and September 2024, so we considered the models available at the time the study was conducted.}. 
    
All models were run with a temperature of 0.2 to reduce the random variability in the output, thus favoring the accuracy and consistency of the generated user stories.

Due to  the company security policies, we used Microsoft Azure as the hosting platform environment for running the experiments with LLMs.

\begin{figure*}[htbp]
    \centering
    \includegraphics[width=\textwidth]{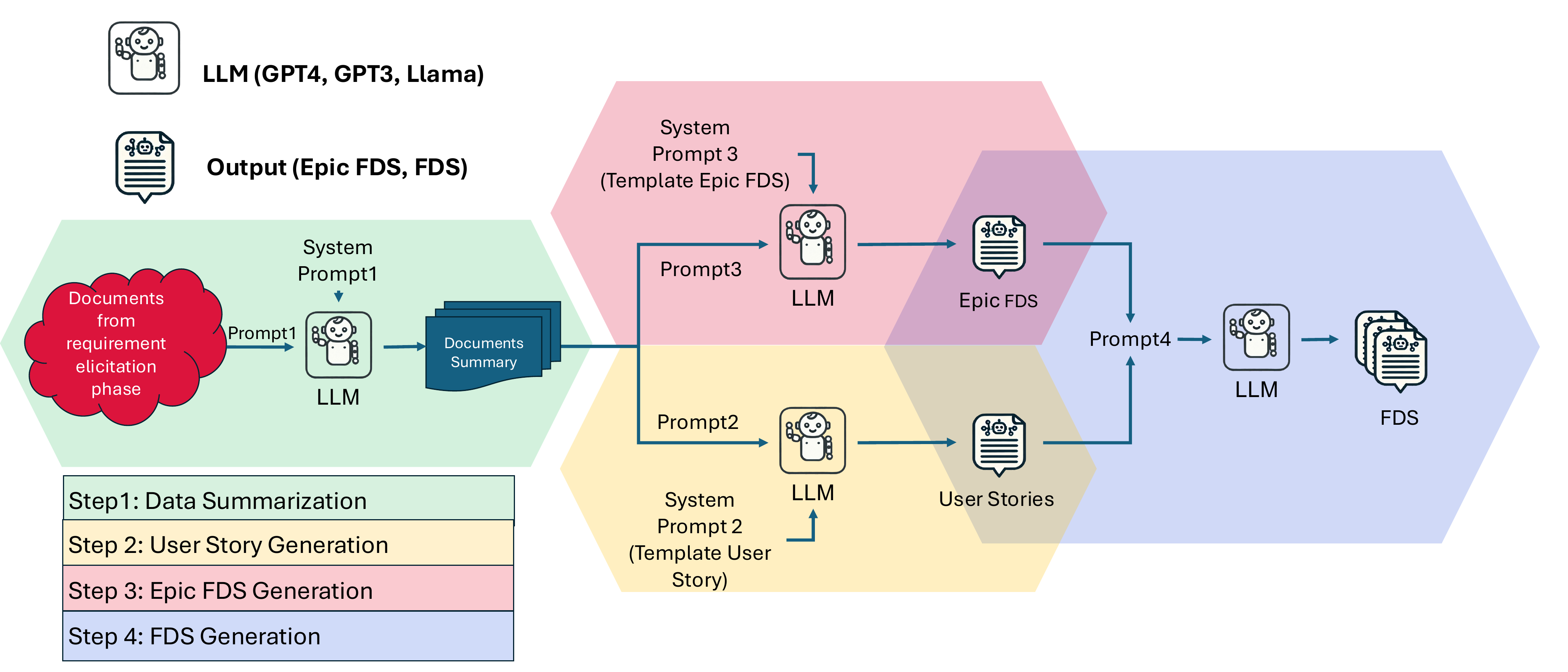}
    \caption{Overview of the LLMs-supported company RE workflow. The process consists of four steps: (1) Data Summarization, (2) User Story Generation, (3) Epic FDS Generation, and (4) LLM-Based FDS Generation.}
    \label{fig:FlowChart}
\end{figure*}

Figure~\ref{fig:FlowChart} illustrates the requirements specification process we mimicked using LLMs, which includes the following steps.

\subsection{Data Summarization}
Since LLMs have a limited token size, they cannot process the entire documents produced by the company during the requirements elicitation phase.
The  objective of this step is to create a concise and focused summary of the most relevant information available in these documents so that LLMs can  use the summary  as the starting point for requirements specification.

\textbf{Input}: 
\begin{itemize}
    \item Documents from the requirements elicitation phase.
\end{itemize}

\textbf{Output}: Documents summary.

The requirement analysts provided the same documents they used to perform the requirement specification phase. These documents include details about availability management, search filters, and discussions about requirements in different contexts, as follows: 

\begin{itemize}
    \item \textbf{Deck 1:} The document is a presentation that provides a high-level view of the strategy for managing media availability. This document contains a description of how the system filters and retrieves relevant data based on parameters defined by the users, together with explanations and examples of how search results are displayed, and performance details like performance metrics. It serves as a foundational guide to creating user stories and functional specifications. 
    \item \textbf{Deck 2:} This document is a presentation delving into the changes and refinements related to search filters and display methods. It includes an explanation of the search filters together with specific examples. Moreover, the presentation shows examples of the various views, including detailed layouts, user-centric presentation, and usability improvements. 
    \item \textbf{Meeting minutes:} The document contains a collection of several meeting minutes derived from "\textit{war room}" meetings\footnote{In this context, 'war room' refers to an intensive, collaborative session used to rapidly elicit and refine system requirements through direct stakeholder interaction.}.
    This document provides contextual and collaborative insights highlighting challenges in aligning requirements across international teams. The document ensures that generated user stories and functional designs account for cross-market differences.
\end{itemize}

In the data summarization process, the LLM\footnote{While we generally use the term LLM without further specification in the text, it is important to note that all three selected LLMs were utilized for each step in the workflow.} summarized the initial documents with the help of a prompt (Prompt1) and a system prompt (System Prompt1) included in the online appendix \cite{appendix2025}.  
The system prompt gives the high-level context to the model, defining persona or \textit{role}, \textit{rules}, and \textit{constraints}. It provides persistent contextual \textit{background} and sets \textit{goals} that govern the model's responses in subsequent interactions.
The prompt is the \textit{task-specific query} provided by the user to the system.
In this step, the \textbf{System prompt 1} was crafted to drive the model to act as an expert assistant (role) in creating detailed summaries.
The purpose of these summaries is to provide the necessary information to generate specific responses during the RE phase (background).
The summaries should focus on the main and most important topics of the attachments, providing a complete and detailed overview of the discussed concepts and proposals (goals). The summaries must exclude dates, links, irrelevant details like ticket codes, locations, countries, and general information about Fincons Group (rules). Finally, the prompt explicitly forbids adding any concluding comments to the generated summary (constraints).
The \textbf{Prompt 1} asks the LLM to summarize the provided documents. 

The prompt design was the outcome of an iterative process conducted by one of the authors in collaboration with a Fincons analyst who possesses experience with the project.

We also checked that the total number of tokens did not exceed the limit allowed by the models\footnote{Token limits for GPT3, GPT4, and Llama are 4096, 32,768, and 4096 respectively \cite{OpenAI2024,LlamaModelCard}.}. 
The output was stored in a single text file for use in the subsequent experimentation phase (see Step 1 of Figure.~\ref{fig:FlowChart}).

\begin{figure*}[htbp]
    \centering
    \includegraphics[scale=0.50]{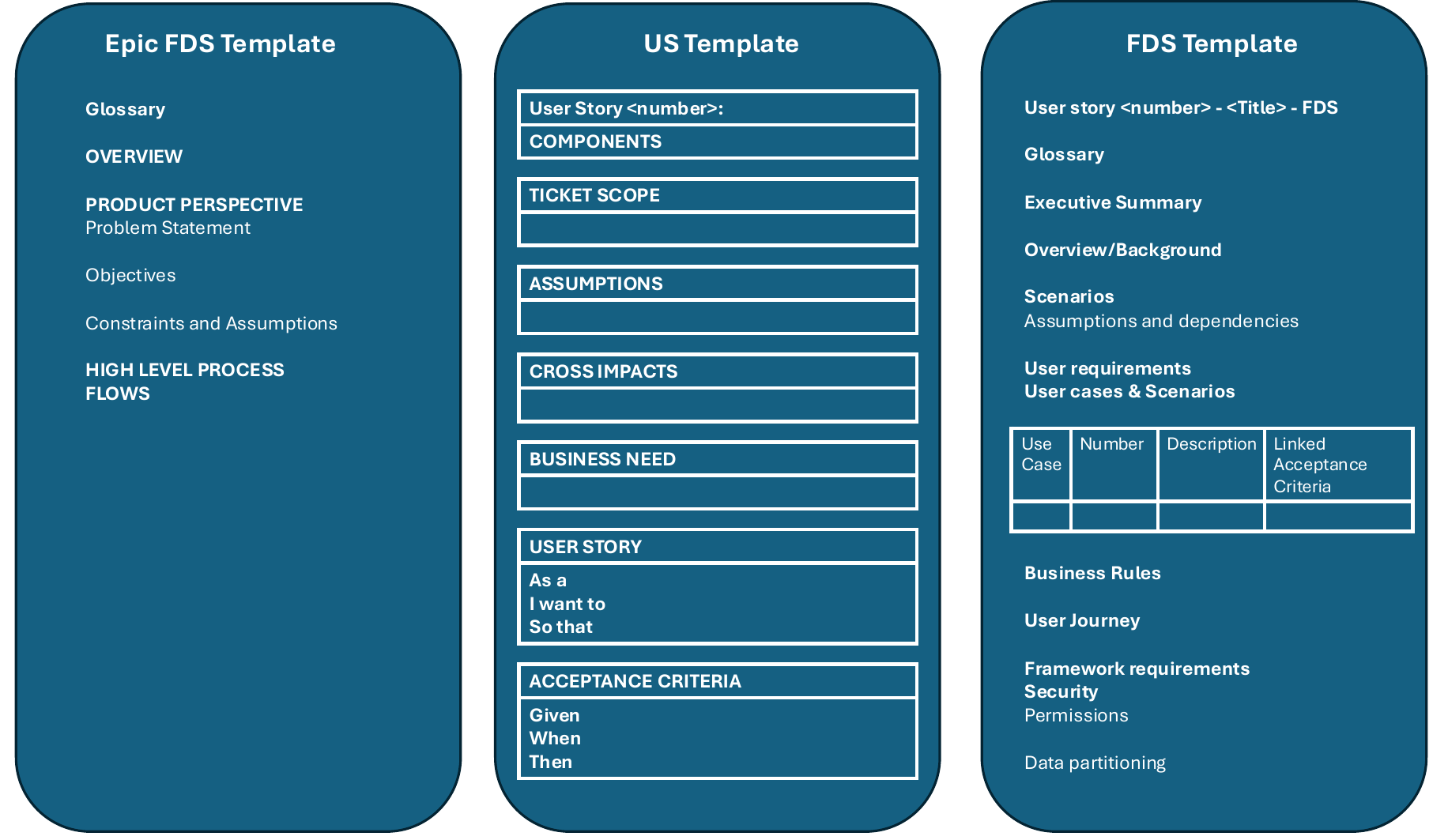}
    \caption{(a) Epic FDS Confluence Template, (b) User Stories Jira Template, (c) FDS Confluence Template}
    \label{fig:templates}
\end{figure*}

\subsection{User Stories generation}
The objective of this phase is to create user stories that comply with the Jira template used by the company (see Step 2 of Figure.~\ref{fig:FlowChart}).\\
\textbf{Inputs}:
\begin{itemize}
    \item Documents summarized in Step 1.
    \item User stories Jira template (short version).
\end{itemize}
\textbf{Output}: User Stories document.

A System Prompt and a Prompt were provided to the LLM (Both System prompts 2 and Prompt 2 are in the Appendix \cite{appendix2025}). 
The \textbf{System prompt 2} starts by establishing the context of the activity, describing \textit{AllRights} as an advanced digital rights management solution, and specifying the main features of the system. Next, the prompt instructs the model to act as a functional analyst in charge of creating a comprehensive set of user stories (US) for the ‘\textit{Availability}’ feature of the \textit{AllRights} project. 
The system prompt includes as an input the US template with the sections (and a description of them) to be completed (Figure \ref{fig:templates}-b).
We note that we prompt the model with a short version of the Jira template used by the company, as we omit from the template all the details that the model cannot be aware of, such as \textit{Status}, \textit{Issue Links}, \textit{Target Release}, \textit{Assignee}, etc\footnote{The interested reader can compare the two versions of the Jira template in the appendix, where for the sake of completeness both the complete and the short version are reported\cite{appendix2025}.}.
\textbf{Prompt 2} asks to generate the user stories covering all possible scenarios, interactions, and system behavior, and requires each user story to be detailed and complete.
Prompt 2 includes as inputs the documents summarized in Step 1 and asks to use the template provided in the system prompt for each user story.

In order to enhance the quality, completeness, and coverage, and identify variation in models, thirty (30) iterations were performed for each model, allowing the exploration of diverse formulations, mitigating the risk of omission, capturing the nuanced aspects, and reducing the influence of outliers. 
Subsequently, the 30 iterations were analyzed: for each US the frequency of occurrences across the dataset and the most frequent combination of components was evaluated to ensure the selected group of USs is representative of the requirements specified in the input documents. This process filters the inconsistent or anomalous results while emphasizing robust recurring outputs.

\subsection{Epic FDS Generation}
The objective of this phase is to generate the Epic FDS document using a template that complies with the Confluence template used by the company (see Step 3 of Figure.~\ref{fig:FlowChart}).\\
\textbf{Inputs}:
\begin{itemize}
    \item Documents summarized in Step 1.
    \item Epic FDS template (short version).
\end{itemize}
\textbf{Output}: Epic FDS document.

The Epic FDS is a critical document that translates business goals into actionable and measurable outcomes, facilitates communication among stakeholders, and ensures consistent delivery of high-quality software features. It serves as a bridge between the business needs expressed in the epic and their technical implementation.

The process begins by crafting System Prompt 3 and Prompt 3 (In online appendix \cite{appendix2025}). \textbf{System Prompt 3} instructs the LLM to assume the role of a functional analyst tasked with creating an Epic FDS document for the "Availability" feature of the AllRights project.
System prompt 3 includes a full description of \textit{AllRights}, explaining its function and main features, and establishing the background necessary for the generation of the document.
The System prompt 3 specifies that the document needs to follow a provided template (Figure \ref{fig:templates}-a) that includes sections for a glossary, overview, product perspective (detailing the problem, objectives, constraints, and assumptions), and high-level process flows, excluding elements such as \textit{Change Control}, \textit{Table of Contents}, and \textit{References}, which contain information inaccessible to the model\footnote{The two versions, short and long, of the Confluence template are in the online appendix\cite{appendix2025}}.
The prompt emphasizes the importance of being comprehensive, omitting no important details, and providing clear and detailed explanations for every point covered, taking into account various events and states affecting availability.
Then, the \textbf{Prompt 3} specifies that the model must produce a complete, detailed, and in-depth Epic FDS document, using the summary produced in Step 1. By processing the prompts and inputs, the model generates the Epic FDS document that will be used in subsequent Step 4.

\subsection{FDS Generation}
The objective of this step is to produce one FDS document for each user story created in the Step 2. The FDS document should be compliant with the company's template in Confluence. \\
\textbf{Inputs}:
\begin{itemize}
    \item USs generated in Step 2.
    \item Epic FDS document generated in Step 3.
\end{itemize} 
\textbf{Output}: One FDS document for each US. 

FDSs accurately describe the functionality and requirements of the system. 

\textbf{System Prompt 4} also begins by outlining the context of the project, describing \textit{AllRights} project, and detailing its main features and functionality.
Next, the prompt instructs the model to act as a functional analyst responsible for creating the FDS documents for each user story associated with the 
\textit{"Availability"} feature of the \textit{AllRights} project.
System prompt 4 specifies that the LLM must follow a comprehensive FDS template that includes sections like a glossary, executive summary, overview, scenarios, assumptions, user requirements, etc. (Figure \ref{fig:templates}-c).
As before, only a short version of the FDS template is provided, omitting information that is beyond the model's knowledge (Short and long templates are in the online appendix \cite{appendix2025}).
 The prompt stresses the importance of thoroughness and detailed explanations within each FDS document to ensure a complete understanding of the functionality being described.

\textbf{Prompt 4} specifies that the model must produce a complete and thorough FDS document for a specific US (the US identifier is specified in the prompt), employing the information provided in the Epic FDS and US documentation generated in Steps 2 and 3 (Both prompts are in \cite{appendix2025}).

\section{Evaluation}\label{sec:eval}

To evaluate the plausibility of the LLM-generated US we assessed whether their content covered those authored by human analysts. We also analyzed the structure and content of the created documents, including the Jira, Epic FDS and FDS templates. Moreover, we evaluated the syntactic and semantic quality of the user stories based on the Quality User Story (QUS) Framework~\cite{Lucassen.RE.2015}.
Finally, an expert analyst reviewed the results and provided insights into the quality of the documentation produced and the perceived usefulness of this technology in their daily work.

\subsection{Content and Structural Analysis}
\begin{figure}
    \centering
    \includegraphics[width=1\linewidth]{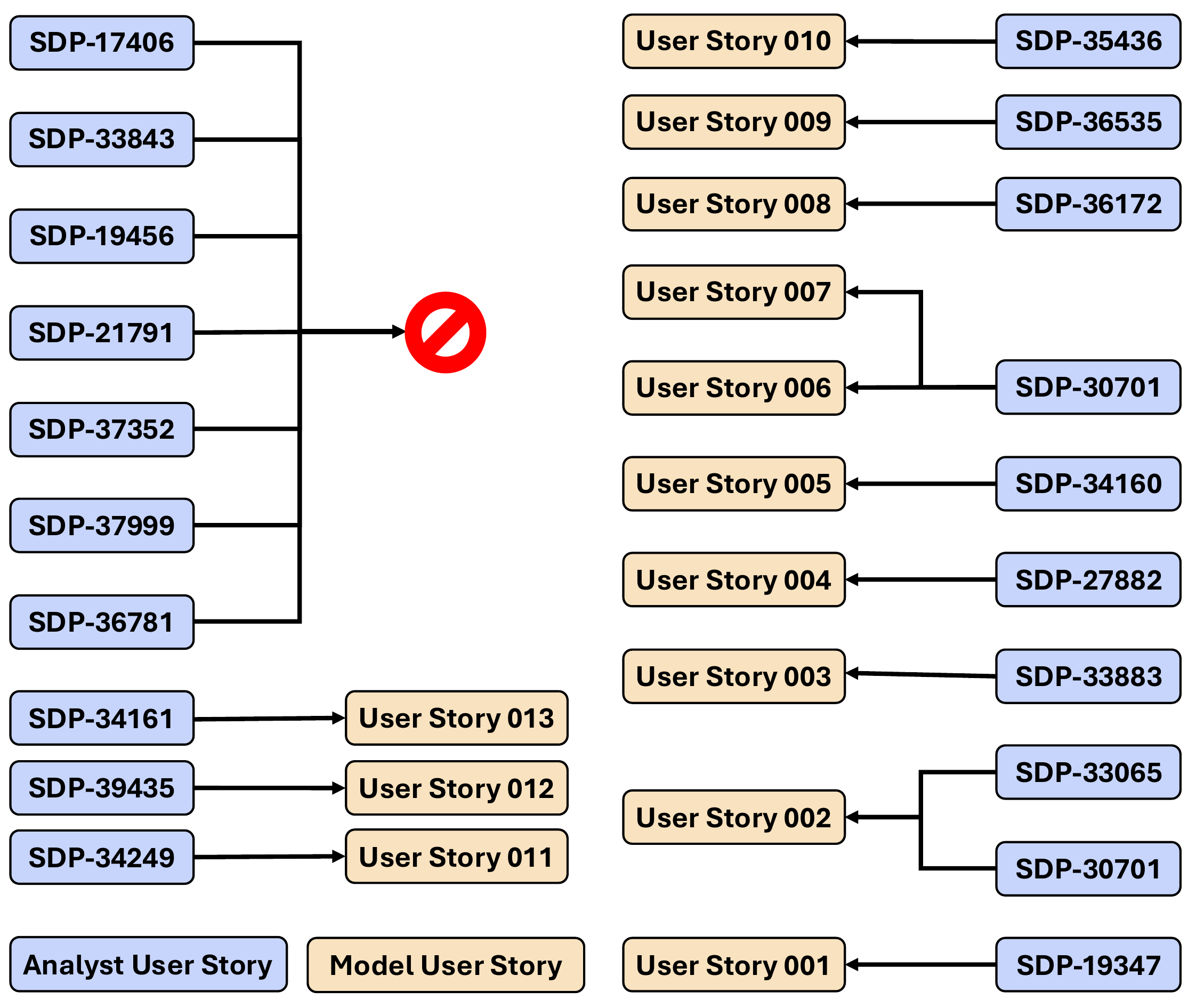}
    \caption{Mapping user stories generated by GPT3}
    \label{fig:Mapping_User_stories_for_GPT3}
\end{figure}

We manually assessed whether one or more LLM-generated US could cover one or more US crafted by the analysts. For the user stories produced by GPT-3 (Figure \ref{fig:Mapping_User_stories_for_GPT3}), we observed instances of many-to-many coverage. For example, a single GPT-3 user story (002) covered the content of two analyst authored user stories (SDP-30701 and SDP-33065). conversely, one analysts' US (SDP-30701) was covered in two separate GPT3-generated US (007 and 006). This indicates situations where US are described at different levels of granularity and specificity.  

\begin{figure}
    \centering
    \includegraphics[width=1\linewidth]{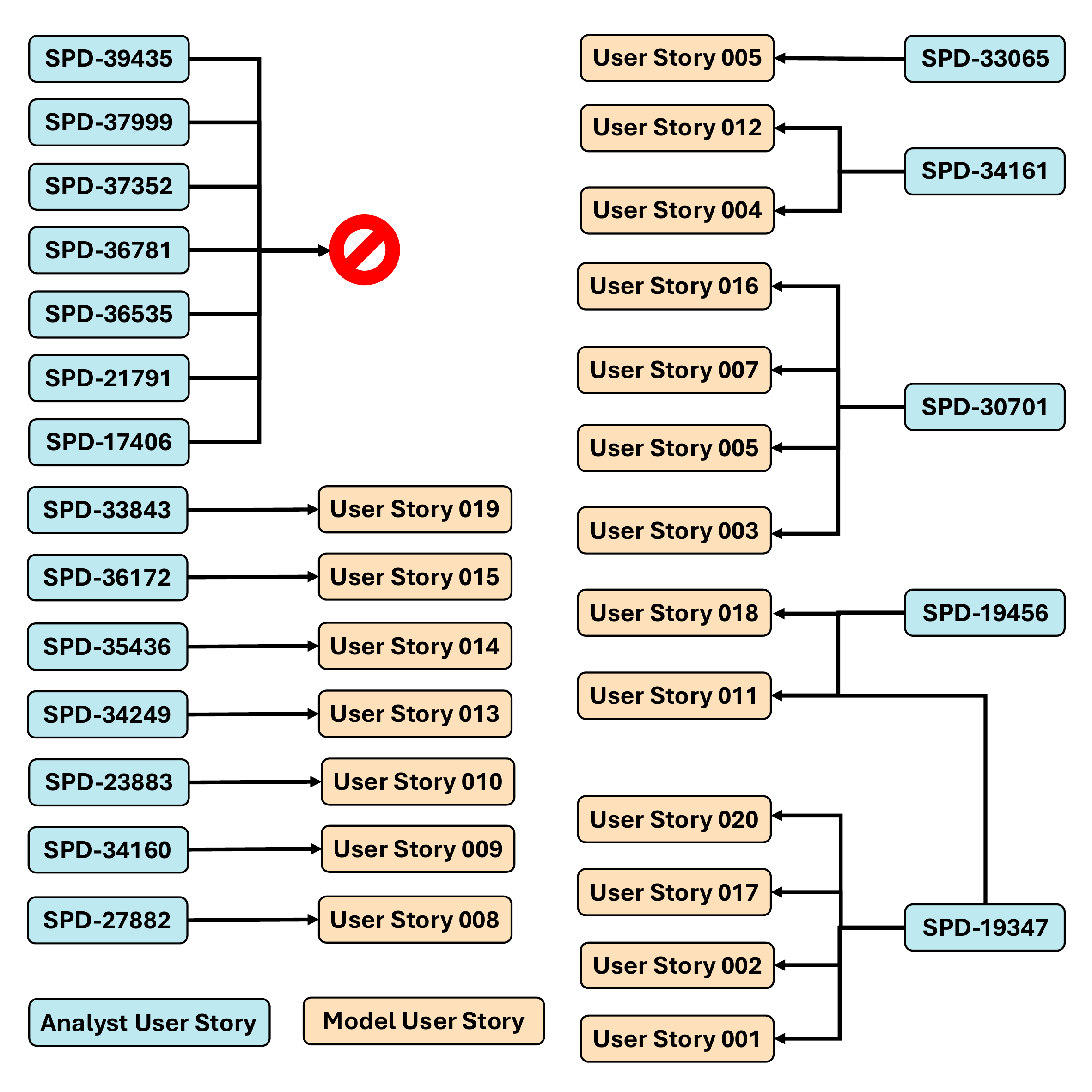}
    \caption{Mapping user stories generated by GPT4}
    \label{fig:Mapping_User_stories_for_GPT4}
\end{figure}

GPT4 consistently specifies more fine-grained US (see Figure~\ref{fig:Mapping_User_stories_for_GPT4}), demonstrating higher specificity than GPT3 in the feature description. 
For example, consider the US SDP-19347: \emph{As a business User, I want to execute a basic search from the availability page so that I can select filters and submit the search
and the system is able to determine if the search will be executed in asynchronous or synchronous way
and based on the check of the system on parameters, the system will approach the standard asynchronous or synchronous process}. 
For this US GPT4 generates 6 user stories, which provide more specificity concerning the search filters and the actor performing the search. For example:
\begin{enumerate}
    \item \emph{As a content manager, I want to search for content availability based on exploitation rights, so that I can identify and utilize available content for distribution.}
    \item \emph{As a sales manager, I want to search for content availability based on sellable rights, so that I can identify and sell available content.}
    \item \emph{...}
\end{enumerate}

Instead, GPT3 generates only one corresponding US which considers one specific actor and includes possible search filters in the same US: 
\emph{As a content manager, I want to be able to search for specific periods, rights values, dimensions, and matrices, so that I can efficiently manage the exploitation and sellable rights for different business entities.}

\begin{figure}
    \centering
    \includegraphics[width=1\linewidth]{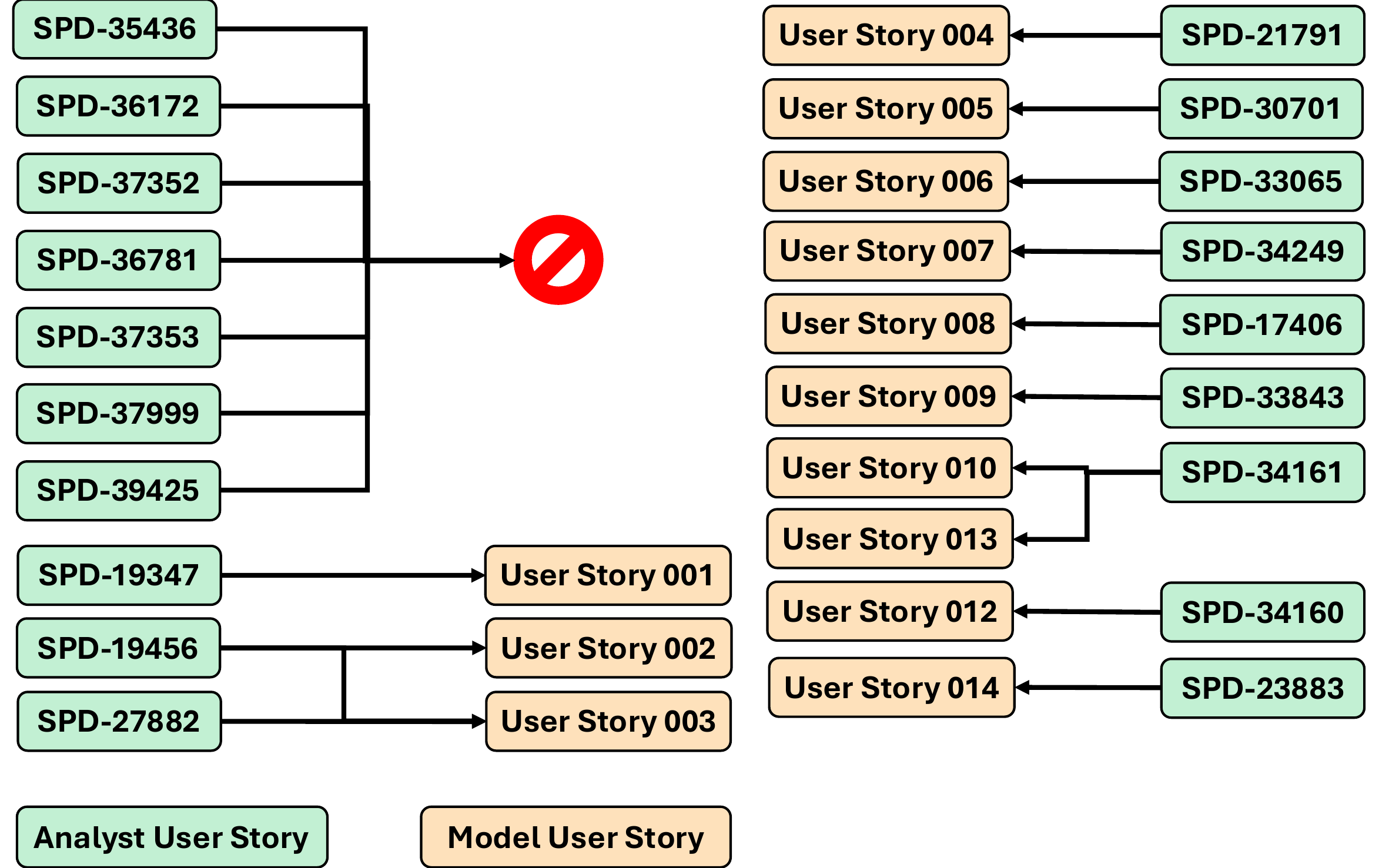}
    \caption{Mapping user stories generated by Llama}
    \label{fig:Mapping_User_stories_for_Llama}
\end{figure}

Almost all US generated by Llama (Figure \ref{fig:Mapping_User_stories_for_Llama})
 have a direct \textit{one-to-one} match with the analyst's US, while some, such as SDP-19456 and SDP-34161, have multiple matches.
This indicates Llama could provide good coverage of the initial requirements while maintaining adequate detail without excessive overlap.

A subset of user stories also remains consistently unaddressed across all models (e.g., SDP-36781, -37352, -37999).
This is due to missing or limited information in the input documents resulting from the requirement elicitation phase. Indeed, knowledge owned by analysts can be tacit and may not always be explicitly provided in the documents resulting from the elicitation phase. 

In other words, if the initial documents do not provide clear and sufficient information, none of the models can generate the corresponding user stories.

We compared the LLMs-generated US and identified essential differences. GPT3 exhibited high redundancy, often producing overlapping or slightly modified versions of similar requirements. GPT4 demonstrated improved contextual understanding, with some user stories mapping across multiple thematic areas rather than being limited to a single category. However, this sometimes led to content fragmentation, where distinct aspects of the same feature were spread across various user stories. Llama created fewer but more concise user stories. While this resulted in structurally cleaner documentation, it also led to gaps in technical specificity. 

Beyond individual user stories, we conducted a comparative analysis of the document generated by the LLMs in Steps 2, 3, and 4. Interested readers can find examples created by GPT-4\footnote{For privacy and confidentiality reasons, we cannot disclose all the generated documents.} in the online appendix \cite{appendix2025}.

\textbf{Jira Template.}  GPT3  generated highly articulated responses, providing acceptance criteria and considerations on system performance. However, the output was sometimes redundant, requiring manual intervention to simplify the requirements. GPT4, on the other hand, produced more concise user stories while maintaining high accuracy and consistency. This balance between detail and synthesis made its user stories easier to interpret. Llama produced more generic and less detailed output, providing a good foundation but lacking the depth necessary for complex projects.

\textbf{Epic FDS Template.} GPT3 produced highly detailed documents, including a lot of additional information, sometimes unnecessary, which complicated readability and required revision.  GPT4 offered a more concise and well-organized output, resulting in a more precise document. The model demonstrated good handling of complexity, providing precise explanations. Llama produced a more straightforward template suitable for contexts where simplicity was preferred but showed shortcomings in complex sections requiring greater depth and organization.

\textbf{FDS Template.} In analysing the FDS templates, GPT3 produced well-structured documents rich in information, providing a clear understanding of the required functionalities but with the risk of overwhelming the reader with redundant information. GPT4, while maintaining a similar structure as GPT3, improved the readability and coherence of the document, presenting information clearly and accessibly. Llama, although technically correct, offered a less detailed template that was less suited for complex documentation, proving more useful in contexts requiring effective summarization rather than detailed analysis.

In summary,  while LLMs enhance efficiency, their outputs often require manual refinement to bridge gaps in completeness and usability.

\subsection{Syntactic and Semantic User Stories Evaluation}\label{aqusa}
We evaluated the quality of the LLM-generated US and compared it with that of the US written by the analysts. For this purpose, we consider the syntactic and semantic qualities suggested in the QUS framework~\cite{Lucassen.RE.2015}.
Syntactic quality is concerned with the textual structure of a US without considering its meaning. Semantic quality is concerned with the relations and meaning of
(parts of) the user story text.

To evaluate \textbf{syntactic quality} we used the \textit{AQUSA} (\emph{Automatic Quality User
Story Artisan})~\cite{Lucassen.REJ.2016}  tool. AQUSA applies a structured set of static checks based on a formal taxonomy of quality criteria, targeting syntactic and structural correctness. The tool outputs detailed metrics capturing the presence and distribution of quality defects.

AQUSA categorizes defects into several distinct types, each representing a different class of quality issues. Among the defects that were present in the US generated by LLMs and analysts we identified:
\begin{itemize}
    \item \textit{conjunctions} (\textit{Conj}): excessive or improper use of conjunctions (e.g., "and") that compromise atomicity.
    \item \textit{uniform} (\textit{Uni}): inconsistencies in the format or structure of acceptance criteria across user stories.
    \item \textit{indicator\_repetition} (\textit{Rep}): redundant specification of role or goal indicators.
\end{itemize}

We report the total number of user stories generated by each model (\textit{Total}), the number of user stories that contained at least one defect (\textit{\# Defected}), and the corresponding proportion of such user stories over the total \textit{\% Def}, as shown in Table \ref{tab:defects-identified}.

\begin{table}[t]
\centering
\footnotesize
\caption{Defects identified in the user stories generated by LLMs and analysts.}
\label{tab:defects-identified}
\begin{tabular}{|l|c|c|c|c|c|c|c|c|}
\hline
    Model & Analysts & GPT-3.5 Turbo & GPT-4 & Llama 3 70B \\ \hline
    \textbf{Total}   & 19 & 13 & 20 & 14 \\
    \textbf{\% Def}   & 37\% & 53.85\% & 10\% & \textbf{7.14\%} \\
    \textbf{\# Def}   & 7 & 7 & 2 & 1 \\
    \textbf{Conj}   & 21.05 & 46.15\% & \textbf{0\%} & 7.14\% \\
    \textbf{Uni}   & 10.53\% & 46.15\% & 10\% & \textbf{0\%} \\
    \textbf{Rep}   & 10.53\% & \textbf{0\%} & \textbf{0\%} & \textbf{0\%} \\
    \hline
    
\end{tabular}
\end{table}

All LLM-generated US do not have the redundancy defect, which instead is present in the US written by the analysts. About half of the US produced by GPT-3 have at least one defect. Overall, GPT-3 produces user stories with syntactically lower quality compared to those authored by the analysts and the other LLM models. In contrast, GPT-4 and Llama produce user stories that have better syntactic quality and are more well-formed compared to those composed by the analysts. Thus, these models could be used to produce US that meet target quality acceptance criteria.

To evaluate the \textbf{semantic quality} of US we  considered the following criteria proposed in the QUS framework. Note that we did not considered quality criteria among sets of user stories because the number of US considered was small. The US generated by the analysts and LLMs satisfied all the quality criteria expressed over the entire set of US.

\begin{itemize}
    \item Feature Specificity (FS): a US should capture a concrete feature.
    \item Rationale Clarity (RC): a US should clearly communicate its rationale, which should not imply a dependency on other system features. 
    \item Problem-Oriented (PO): a US should only specify the
problem, not the solution.
\item Language Clarity (LC): a US should not be specified using ambiguous language.
\item 
    Internal Consistency (IC):  a US should not contradict itself causing inconsistencies.
\end{itemize}

Two authors scored each criterion on a 3-point scale: 1 (non-acceptable), 2 (acceptable), or 3 (good). After the scoring phase, they met to resolve conflicts and agree on a uniform scoring. Based on these scores, we computed two aggregate measures for each model (Table \ref{tab:all-in-one}):

\begin{itemize} 
\item \textit{Avg Score (Avg)} – the average of all quality ratings assigned across all criteria and US from a given source. This provides an overall indicator of the typical quality level observed. 
\item \textit{Rejection Rate (RR)} – the percentage of US that received a score of 1 (non-acceptable) on at least one of the five quality criteria. This metric reflects how many US fail to meet minimal quality expectations. 
\end{itemize}

\begin{table}[t]
    \caption{Semantic quality of USs generated by LLMs  and analysts.}
    \label{tab:all-in-one}
    \centering
    \begin{tabular}{ccccc}
    \hline
    \textbf{Model} & Analysts & GPT-3.5 Turbo & GPT-4 & Llama 3 70B \\ 
    \hline
    \textbf{FS} & 2.38 & 2.31 & \textbf{2.4} & 2.21 \\
    \textbf{RC} & 2.92 & 2.85 & \textbf{2.95} & 2.71 \\
    \textbf{PO} & 2.85 & 2.85 & \textbf{2.9} & 2.86 \\
    \textbf{LC} & \textbf{2.62} & 2.46 & 2.4 & 2.36 \\
    \textbf{IC} & 2.77 & \textbf{3} & 2.95 & 2.86 \\
    \textbf{Avg} & 2.71 & \textbf{3} & 2.95 & 2.86 \\
    \textbf{RR} & \textbf{15.79\%} & 23.08\% & 20\% & 28.57\% \\
    \hline
    \end{tabular}
\end{table}

GPT4 performs better than the other models for all semantic criteria except for language clarity (LC), indicating that the US written by the analysts are less ambiguous. GPT3 and Llama perform worse or the same as  the human analysts except for the internal consistency where GPT3 has the best performance. It also worth noticing that the US generated by the analysts obtained the smallest rejection rate meaning that they can still produce more user stories passing the acceptance criteria.

\subsection{Interview with a Requirement Analyst}
To gain more insights into the results produced by the LLM and to also collect qualitative data, we interviewed the analyst who originally created the Epic FDS, US, and FDS documents.
Due to the analyst's limited time and availability, we asked analysts to assess the quality of each category by reviewing one representative of the group. Here, representative refers to the document that best reflects the typical structure and content generated by the model, selected based on the most frequent number of items (mode) and the most recurring combination of components across all generations, ensuring it exemplifies the model’s average behavior.
The interview included open and closed-ended questions which we reported together with the interview transcripts in the online appendix~\cite{appendix2025}.

GPT4 achieved the highest readability score, while GPT3 had more redundant information. The analyst noted that LLM-generated stories were coherent but often required additional refinement to match human-created ones. For example, they mentioned that "\emph{although coherent the US provide a list of feature without giving specificity on the concrete functionality to be implemented".}
While the produced Epic FDS followed the structured template, the content often resembled a high-level checklist rather than a comprehensive document, requiring significant analyst input.
The analyst noted that the FDS document lacked technical depth and precision, making it difficult to use it directly in software development. The analyst found that a human-written FDS document was more contextually accurate and detailed.

To evaluate the acceptance and perceived usability of our approach, we adopted the Technology Acceptance Model (TAM)~\cite{4af0105e-77c6-3217-8650-3ef6ef5e917a}. This is a widely accepted framework in evaluating new technologies \cite{10.3897/jucs.104938} and has been applied across domains, including software engineering, educational technology, and intelligent systems.

Table~\ref{tab:tam_evaluation} summarizes the TAM answers, revealing that while LLMs improved efficiency and provided useful support requirements specification, the generated FDS still required manual intervention for refinement and adaptation. The analyst considered LLM-assisted documentation  helpful in the initial drafting phase but unreliable for standalone use.

\begin{table}[h]
    \caption{TAM Results}
    \label{tab:tam_evaluation}
    \centering
    \begin{tabular}{
    >{\centering\arraybackslash}p{0.30\linewidth} 
    >{\centering\arraybackslash}p{0.15\linewidth} 
    >{\raggedright\arraybackslash}p{0.40\linewidth}
    }
    \hline
    \textbf{Evaluation Criteria} & \textbf{Score} & \textbf{Notes} \\
    \hline
    LLMs help complete tasks faster & 3/5 & AI was useful in initial drafting but required refinement. \\
    LLMs improve document quality & 3/5 & Quality was inconsistent, requiring post-processing. \\
    LLMs increase productivity & 4/5 & Helped streamline documentation but not without oversight. \\
    LLMs provide useful support in RE & 4/5 & Effective for assisting in user story drafting. \\
    LLMs are useful in daily work & 4/5 & Analysts found AI-generated drafts beneficial. \\
    \hline
    \end{tabular}

\end{table}

The interview revealed that LLMs effectively generate coherent and well-structured documents, such as user stories and epics, but lack detail and adaptability. While user stories were clear and concise, they were incomplete, and epics were simplistic outlines lacking full functionality coverage. The technical specifications created were also not detailed enough for effective development guidance. Despite these limitations, LLMs can provide a time-saving benefit as the analyst will not start from "a blank sheet", and can also rely on a summary of documents. According to the analyst, drafting these documents could result in time savings of 10\% to 15\%. 
Additionally, TAM assessment showed a moderately positive perception of LLMs' usefulness, especially for repetitive tasks and standardized documentation. Overall, LLMs can enhance productivity in the early phases of documentation and definition of requirements but cannot replace human intervention. Expert analysts are still needed to ensure the technical accuracy, completeness, and adaptability of documents, as LLM-generated drafts require revision and integration for complex projects.

\label{sec:results}

\section{Discussion of Findings}
\label{sec:discussion}

The integration of LLM in the requirements specification phase can bring several advantages.

\textbf{Efficiency gain}. LLMs can reduce the time needed for the initial documentation phase. If document templates are provided and no tacit knowledge is present in the elicitation documents, LLMs can produce detailed user stories, Epic FDSs and FDSs.
This time saving offers potential productivity gains in early RE phases.

\textbf{Quality evaluation}.
From all the different evaluations, GPT4 emerges as the best model with the soundest contextual understanding, even if sometimes it leads to content fragmentation, where distinct aspects of the same feature are spread across various user stories.
GPT4 excelled at generating concise, accurate, and consistent user stories that were easier to interpret compared to other methods. It produced a more precise and well-organized Epic FDS document with improved readability. Notably, LLMs, including GPT4 and Llama, avoided the redundancy found in analyst-written user stories and demonstrated better syntactical quality. While LLMs show promise for generating user stories that meet semantic quality standards, analyst-written user stories were found to be less ambiguous in terms of language clarity compared to GPT4. LLMs can be used to improve the syntactical quality and structure of user stories.

\textbf{Lack of depth and precision and need for human oversight}.
The evaluation confirms that despite their structural adherence to templates, LLM-generated documents often require human post-processing to meet technical and domain-specific requirements. 

The analyst, who evaluated the results, estimates that the time required for drafting these documents can be reduced by 10\% to 15\% and that the documents produced require additional refinements, being sometimes too high-level and lacking technical depth and precision. Nevertheless, the answer of the analyst could be biased~\cite{10.1145/3544548.3581025}. Examples are \textit{overconfidence bias}~\cite{articleIdentityThreats} (overestimating their own abilities and the unique complexity of their work, while believing they are unreplaceable), \textit{confirmation bias} \cite{BASHKIROVA2024100066} (assuming that an AI-based system can make mistakes, and ignoring or minimizing its capabilities), and \textit{status quo bias} \cite{10.3389/fdgth.2022.966174} (an inclination to reject the possibility of using new AI-based technology by downplaying its feasibility to maintain the status quo). To mitigate these biases, we balanced the analyst evaluation with the syntactic and semantic evaluation of US quality suggested in the QUS framework. 
Furthermore, we would like to point out that we involved only one analyst in the interview, as this analyst is the original author of the FDS and is well-versed in the entire project. She was the only one in the company with all the necessary information to assess the LLM-generated documents.

\textbf{Input Dependency}.
We found that certain user stories were consistently missed by all models, due to insufficient information in the requirements elicitation documents. This is because analyst knowledge is often tacit and not fully captured.
However, this limitation can be mitigated by integrating LLMs into the requirements elicitation phase. For example, LLMs can serve as automated 'note-takers' to enhance documentation and also identify parts in the requirements elicitation documents where ambiguities and tacit knowledge may be present, to trigger improvements from the analysts.

\textbf{Cost Analysis}.
Additionally, we conducted a cost analysis to evaluate the models in financial terms.
Table \ref{tab:modelCostsTable} shows the cost per 1,000 tokens, total token usage for each model and overall expenditure on cloud computing resources. The result revealed that GPT4 was the most expensive model primarily due to its high output cost and extensive token generation. 
On the other side, GPT3 was the cheapest model, but its redundancy meant that token efficiency was lower, requiring additional human revision. Llama fell in the middle, offering a cheaper alternative than GPT4 while still providing structured outputs.
In contrast, GPT3 and Llama offer more economical alternatives, though their outputs require additional refinement.

\begin{table}[h]
    \caption{Comparison of Input and Output Costs for Different Models (Costs are €/1,000 tokens)}
    \label{tab:modelCostsTable}
    \centering
    \resizebox{\columnwidth}{!}{
    \begin{tabular}{lccc}
        \toprule
        \textbf{Model} & \textbf{Input Cost} & \textbf{Output Cost} & \textbf{Total Model's Cost} \\ 
        \midrule
        GPT3 & 0.0010 & 0.0019 & 3.60€  \\
        GPT4 (128k context) & 0.0047 & 0.0141 & 24.80€\\
        Llama & 0.0034 & 0.0099 & 11.80€\\
        \bottomrule
    \end{tabular}}
\end{table}

Our approach has several limitations discussed below.

\textbf{Data and domain specificity.} This study was conducted in collaboration with an industrial partner, focusing on a specific project. Consequently, the generalizability of these findings to other domains or projects necessitates further investigation. Nevertheless, following this initial experimentation, the company has decided to expand its exploration of LLMs in other requirements engineering tasks, such as the identification of  Acceptance Criteria (AC). This study has instilled confidence in them to pursue continued research on these tools in partnership with the academic community.

\textbf{Token limits.} A major challenge during experimentation was the limitation of token context windows, particularly for GPT3 and Llama-3, which have smaller input capacities compared to GPT4. To address this limitation, raw input data (e.g., documentation, meeting notes, presentations) had to be summarized manually before feeding it into the models. This summarization not only added an extra processing step, but also increased the risk of critical information loss, leading to less efficiency for the models. In this case, GPT4’s extended token capacity made it well-suited for handling long technical documents, reducing the need for excessive data condensation.
However, this limitation can be overcome as several new models have already been released that boast context windows far exceeding even the earlier versions of GPT4 used in this experiment\footnote{As a way of example, Google Gemini 1.5 Pro has a 1 million token context window.}. 
  
\textbf{Model dependency.} We acknowledge that our experiment is heavily dependent on the LLMs chosen and that the results could be substantially different using a different or newer model. However, we aimed to understand not only if LLMs could produce high-quality results (which was the case of GPT4 in particular), but also if analysts could easily integrate them into their daily routine. As we can see, there is still a good dose of skepticism and room for improvement. This disparity can be effectively addressed if companies continue to conduct controlled experiments, such as the one presented in this paper, rather than relying solely on individual employee initiatives.

\section{Conclusion}\label{sec:conclusion}
Overall, the study demonstrates that while LLMs have the potential to revolutionize RE by automating repetitive documentation tasks and enhancing early-phase productivity, they are not yet ready to replace human expertise entirely. 
The findings advocate for a synergistic approach where LLMs serve as effective drafting tools while human analysts provide the critical contextual and technical oversight necessary for high-quality requirements documentation. 
In the future, we will continue to investigate with the company ways of optimizing synergic collaboration between humans and the AI to further enhance the efficiency and quality of RE tasks.

\section{Acknowledgment}
TITANUX - “creaTing hIgh qualiTy ANd sEcure systems that emphasize UX”, a project funded by the Apulia Region under the Call for Tenders “General Regulation of Exempted Aid Schemes” No. 9 of 26/06/2008 and subsequent amendments and additions - Title IX “Aid to Small Enterprises for Integrated Facilitation Projects”; Project Code: Z85MUK8.


\end{document}